\documentclass[12pt]{amsart}
\usepackage{hyperref}

\usepackage{xcolor}

\newcommand{\ndash}{\nobreakdash-\hspace{0pt}}

\newcommand{\ii}{{\mathrm{i}}}
\newcommand{\dd}{{\mathrm{d}}} 
 
\newcommand{\EE}{{\mathrm{e}}}

\DeclareMathOperator{\Map}{Map}

\newcommand{\ev}{\mathrm{ev}}

\newtheorem{Thm}{Theorem}[section]
\newtheorem{Prop}[Thm]{Proposition}
\newtheorem{Lem}[Thm]{Lemma}

\newtheorem*{Thm*}{Theorem}
\newtheorem*{Lem*}{Lemma}

\theoremstyle{remark}

\theoremstyle{definition}

\newtheorem{rmk}[Thm]{Remark}

\newcommand{\C}{\mathsf{C}}

\newcommand{\calA}{\mathcal{A}}

\newcommand{\calC}{\mathcal{C}}

\newcommand{\calI}{\mathcal{I}}
\newcommand{\calO}{\mathcal{O}}

\def\gpd{\,\lower1pt\hbox{$\longrightarrow$}\hskip-.24in\raise2pt
               \hbox{$\longrightarrow$}\,}

\newcommand\es{\EE^{\frac\ii\hbar S}}
\newcommand\ems{\EE^{-\frac\ii\hbar S}}

\newcommand{\calL}{{\mathcal{L}}}
\newcommand{\calN}{{\mathcal{N}}}
\newcommand{\DeltaS}{{\Delta_S}}
\newcommand{\DeltaST}{{\Delta_{S,T}}}

\newcommand{\bea}{\begin{eqnarray}}
\newcommand{\eea}{\end{eqnarray}}
\newcommand{\nn}{\nonumber}

\def\F{\mathcal F}

\def\g{\mathfrak g}
\def\c{\mathbf c}
\def\superPhi{\mathbf \Phi}
\def\superXi{\mathbf \Xi}
\def\L{{\cal L}}

\def\ev{{\mathrm{ev}}}

\def\M{{\mathcal M}}
\def\Map{{\mathrm{Map}}}

\def\dW{{d_W}}
\def\A{{\mathcal A}}
\def\Der{{\rm Der}}
\def\C{{\mathbb C}}
\def\k{{\mathfrak k}}
\def\Cu{{\mathcal C}}
\def\I{{\mathcal I}}
\def\L{{\mathcal L}}

\def\beq{\begin{equation}\begin{aligned}}
\def\eeq{\end{aligned}\end{equation}}

\begin{document}

\title{Equivariant Batalin-Vilkovisky formalism}

\author{F.~Bonechi}\address{INFN Sezione di Firenze, Via G. Sansone 1, 50019 Sesto Fiorentino, 
Firenze, Italy}\email{francesco.bonechi@fi.infn.it}

\author[A.~S.~Cattaneo]{A.~S.~Cattaneo}
\address{Institut f\"ur Mathematik, Universit\"at Z\"urich\\
Winterthurerstrasse 190, CH-8057 Z\"urich, Switzerland}  
\email{cattaneo@math.uzh.ch}

\author{J.~Qiu}\address{Department of Mathematics, Uppsala University, Box 480, 75106 Uppsala, Sweden}\email{jian.qiu@math.uu.se}

\author{M.~Zabzine}\address{Department of Physics and Astronomy, Uppsala University, Box 516, 75120 Uppsala, Sweden}\email{maxim.zabzine@physics.uu.se}


\date{}

\maketitle

\begin{abstract}
We study an equivariant extension of the Batalin-Vilkovisky formalism for quantizing gauge theories. Namely, we introduce a general framework to encompass failures of the quantum master equation, and we apply it to the natural equivariant extension of AKSZ solutions of the classical master equation (CME). As examples of the construction, we recover the equivariant extension of supersymmetric Yang-Mills in 2d and of Donaldson-Witten theory.
\end{abstract}

%


\section{Introduction}

One cannot overlook the role played by equivariant methods in quantum field theory in the last thirty years.
 Different versions of equivariant localization played central role in obtaining the exact results for supersymmetric 
 field theories. One prominent example is the construction of  $N=2$ 4d supersymmetric theory in 
  $\Omega$-background \cite{nekrasov03}  (the equivariant version of the Donaldson-Witten theory).
   Later these ideas were implemented and generalized to other field theoretical examples, mainly within supersymmetric 
    field theory context. 
         In any calculation using supersymmetric localisation for supersymmetric gauge theories, it is inevitable that the BRST transformation is combined with supersymmetry. The BRST transformation, which is a special case of the BV algebra, squares to zero, while the supersymmetry transformations square to infinitesimal actions of isometries (plus possible global symmetries). From this point of view, there seems to be some tension between the two. The problem is further compounded by the presence of ghost zero modes which forces us to treat the constant gauge transformation also equivariantly. This means that the BRST transformation will now also square to constant gauge transformation, and must be treated as equivariant differential.
    Here we would like to address the equivariance from a more   systematic gauge theoretical point of view, 
    namely within the the Batalin-Vilkovisky formalism. 

Many of the above examples of gauge theories whose equivariant extension proved to be so fruitful have a very simple description in terms of the Batalin-Vilkovisky (BV) 
formalism and in particular can be formulated as AKSZ actions \cite{AKSZ:geometry_of_BV}  (also see \cite{roytenberg:courant2} for an introduction).
 Namely, in order to construct the BV extension of a given action one has to double the fields and the ghosts by adding the antifields;
   in this way one gets an odd symplectic manifold with induced odd Poisson bracket denoted by $\{ , \}$\footnote{In the literature, the notation $(\ ,\ )$ is also common.} and the BV action is a degree zero solution of the {\it classical master equation} (CME) \cite{BV1, BV2, BV3}
\begin{equation}
 \label{CME}
 \{S,S\}=0\;,
\end{equation}
such that the original gauge invariant action is recovered from $S$ by putting the antifields to zero. Provided $S$ is extended to solve the {\it quantum master equation} 
(\ref{QME}), the path integral of the gauge fixed theory is then recovered by integrating $\exp ( i S/\hbar )$ over a Lagrangian submanifold obtained by fixing the antifields in a way that the action is now nondegenerate; invariance under the change of gauge fixing is interpreted as invariance of the BV integration under the deformation of the Lagrangian submanifold. 

The AKSZ construction provides a solution $S$ of the CME (\ref{CME}) in terms of a very trasparent geometrical procedure. 
Indeed, it is very easy extend the AKSZ solution to an action satisfying the equivariant version (\ref{modCME}) of (\ref{CME}). On the 
other hand, since the CME is not anymore satisfied, the BV formalism must be modified, in particular one has to understand how to guarantee the
invariance of the path integral under the change of Lagrangian submanifold. This paper is devoted to developing the proper setting to deal with equivariance in the 
framework of the Batalin-Vilkovisky method.  Our main goal is to suggest the systematic derivation of the 
equivariant odd transformations (the mixture of supersymmetry and BRST symmetry)
 which appear in the localization literature in a rather ad hoc fashion.  Moreover, we hope that in the future the suggested framework can be  extended to manifolds with boundaries (in analogy with
  \cite{Cattaneo:2012qu})
  and that we can understand better the structural questions behind the ''localizable'' theories.

In Section \ref{s:QME} 
 we discuss how we can encompass actions that fail to solve the quantum master equtions still keeping the spirit of the BV formalism, {\it i.e.} invariance 
of the integral under deformations of the Lagrangian submanifold. This is in principle possible provided we accordingly restrict the class of observables and of Lagrangian submanifolds in a 
way that is compatible with the failure $T$ of the QME. Apart from some additional conditions, this setting is equivalent to working in the symplectic reduction defined by the zero locus 
$\calC_T$ of $T$ (see Remark \ref{rmk_coiso_red}). In Section \ref{s:AKSZ}
  we apply this formalism to the equivariantly extended AKSZ solution. The formalism leads us  to 
consider a complex that is a quantum version of the Cartan model for the equivariant cohomology of the $\g$-differential algebra defined on the 
space of AKSZ fields. In Section \ref{s:2D} we consider as an example SUSY Yang-Mills in two dimensions. The equivariant extension was considered in \cite{nekrasov09}. Here we prove 
that BV complex of fields contains the supersymmetric multiplet in the $\Omega$-background considered in \cite{nekrasov09}. Moreover, we study the equivariant observables by using a method 
that was introduced in 
\cite{bciz:EquivA-mod}. In Section \ref{s:4D} we study the AKSZ version of the topological twist of $N=2$ supersymmetric
4d Yang-Mills theory considered in \cite{w88} and its equivariant extension \cite{nekrasov03, nkerasov-ok}. 

\subsubsection*{Aknowledgements}

 We thank the anonymous referee for the valuable comments.
A. S. C. acknowledges partial support of SNF Grant No. 200020- 172498/1. This research was (partly) supported by the NCCR SwissMAP, funded by the Swiss National Science Foundation. M. Z.  acknowledges the partial support by Vetenskapsr\r{a}det under grant \#2014-5517, by the STINT grant, 
and by the grant  ``Geometry and Physics"  from the Knut and Alice Wallenberg foundation.

\bigskip
\bigskip

\section{Relaxing the Quantum Master Equation}\label{s:QME}

The fundamental fact of the BV formalism is that, given a family $\calL_t$ of Lagrangian submanifolds of the BV space and a half density $\rho$, one has
\[
\frac\dd{\dd t}\int_{\calL_t}\rho = 0,
\]
if $\Delta \rho =0$ with $\Delta$ being the canonical BV Laplacian on half densities\footnote{The discussion in this section is properly rigorous in finitely many dimensions. The canonical BV Laplacian on half densities was introduced by Khudaverdian in \cite{Khudaverdian:1989si} and cannot be extended to the infinite dimensional context. To apply this discussion to field theory we have to assume a regularization as usual.}.

Typically we fix a reference $\Delta$\ndash closed half density $\rho$ and on the algebra $\calA$ of functions
we define 
$\Delta f:=(\Delta(\rho f))/\rho$, for any $f\in\calA$, where in the r.h.s.\ we use the canonical BV Laplacian on half densities.
By $\int_\calL f $ from now on we mean $\int_\calL f \rho$. The above statement now becomes
\[
\frac\dd{\dd t}\int_{\calL_t}f = 0,
\]
if $\Delta f =0$.

The main application of this is that the integral of a $\Delta$\ndash closed function is invariant under deformations of the Lagrangian submanifold on which we integrate. 
More generally, the integral of $f$ is invariant under deformations if we restrict ourselves to the class of Lagrangian submanifolds on which $\Delta f$ vanishes.
 We will pursue this idea here. 

In quantum field theory, one usually considers functions of the form $\es$, where $S$ is a function of even degree. From the properties of the BV Laplacian on functions, 
it follows that $\Delta\es=0$ if and only if $S$ satisfies the quantum master equation
\begin{equation}\label{QME}
\frac12\{S,S\}-\ii\hbar\Delta S=0.
\end{equation}
In this case, the ``gauge fixed partition function" $\int_\calL\es$ is invariant under deformations of $\calL$. One is also interested in inserting a second function 
$\calO$, called a preobservable, in the integral. One then has that also $\int_\calL\es\calO$ is invariant under deformations of $\calL$ if, in addition, 
$\DeltaS\calO=0$,  where $\Delta_S$ is the $\es$ twisted
coboundary operator defined as
\[
\DeltaS\calO:=\ems\Delta(\es\calO)=\Delta\calO+\frac\ii\hbar Q\calO
\]
with $Q:=\{S,\ \}$. One calls a $\Delta_S$\ndash closed preobservable an observable.

More generally, without assuming the quantum master equation, we define
\begin{equation}\label{qme_anomaly}
T:=\left(\frac\hbar\ii\right)^2\ems\Delta\es=\frac12\{S,S\}-\ii\hbar\Delta S
\end{equation}
and note that now
\begin{equation}\label{twisted_BV}
\DeltaST\calO:=\ems\Delta(\es\calO)=\Delta\calO+\frac\ii\hbar Q\calO+\left(\frac\ii\hbar\right)^2T\calO.
\end{equation}
In particular, since $T$ is proportional to $\DeltaST 1$ we get $\DeltaST T=0$, which, using the fact that $T$ is odd and hence satisfies $T^2=0$, gives
\begin{equation}\label{e:DeltaT}
\Delta T+\frac\ii\hbar QT=0.
\end{equation}
As remarked above, $\int_\calL\es$ is invariant under deformations of $\calL$ if we restrict ourselves to the class of Lagrangian submanifolds on which $T$ vanishes. 
We will call them  $T$-Lagrangian submanifolds and from now on we restrict our attention to this class of 
 Lagrangian submanifolds.  We then observe the following:
\begin{enumerate}
\item $\int_\calL\es\calO=0$ if $\calO$ is proportional\footnote{The proportionality is defined by some
 functional, not necessarily constant. As indicated below, ``proportional to $T$'' means in the ideal $\calI_T$ generated by $T$.} to $T$, and
\item $\int_\calL\es\calO$ is invariant under deformations of $\calL$ if $\DeltaST\calO$ is proportional to $T$.
\end{enumerate}
This suggests working modulo the ideal $\calI_T$ generated by $T$. Note however that
\begin{equation}\label{DeltaSTderivation}
\DeltaST(T\calO)=-T\DeltaST\calO-\{T,\calO\}.
\end{equation}
This means that $\calI_T$ becomes a $\DeltaST$\ndash differential ideal only after restricting to the subalgebra $\calN_T$ of functions that 
Poisson commute with $T$, possibly up to a term proportional to $T$:
\begin{equation}
 \label{NT}
 \calN_T = \{\calO\in\calA \ |\ \{T,\calO\}\in \calI_T\}\;.
\end{equation}

Note that $T$ is contained in $\calN_T$,
since by degree reasons $\{T,T\}=0$. Actually,
$\calN_T$ is the Lie normalizer of $\calI_T$ (i.e., the largest Lie subalgebra of $(\calA,\{\ ,\ \})$ that contains $\calI_T$ as a
Lie ideal). As a consequence, $\calA_T:=\calN_T/\calI_T$ inherits the structure of a Poisson algebra, whose elements we call the {\it quantum preobservables}. Moreover, $\DeltaST$ 
descends to a coboundary operator on $\calA_T$ by
\[
\DeltaST[\calO]:=[\DeltaST\calO]=\left[\Delta\calO+\frac\ii\hbar Q\calO\right].
\]
We call a $\DeltaST$\ndash closed quantum preobservable a {\it quantum observable}. Note in particular that the unit $1$ belongs to $\calN_T$ and
that its equivalence class $[1]$ is a unit in $\calA_T$ and an observable.

We may finally summarize the above discussion by observing that
\begin{enumerate}
\item for every observable $[\calO]$ we may define $\int_\calL\es[\calO]$ as $\int_\calL\es\calO$ where $\calO$ is any representative in $[\calO]$, and
\item $\int_\calL\es[\calO]$ is invariant under deformations of $T$-Lagrangian $\calL$ if $[\calO]$ is a quantum observable.
\end{enumerate}

\begin{rmk}\label{rmk_coiso_red}
The Poisson algebra $\calA_T$ may also be interpreted as the algebra of $\{T,\ \}$\ndash invariant elements in
$\calA/\calI_T$, which in turn may be interpreted as the algebra of functions on the zero locus $\calC_T$ of $T$. Thus, we 
may interpret $\calA_T$ as the algebra of functions on the symplectic reduction $\underline\calC_T$ of $\calC_T$.
Moreover, the condition that $T$ vanishes on a Lagrangian submanifold $\calL$ geometrically means that $\calL$ is contained in $\calC_T$. We may then be tempted to 
interpret the whole theory as the usual BV formalism but on $\underline\calC_T$. This is correct if $\es$ is in $\calN_T$. Notice however that a gauge fixing Lagrangian submanifold
contained in $\calC_T$ necessarily contains the characteristic foliation generated by $\{T,-\}$ so that this cannot be a full gauge fixing. For this reason we have to assume
that the leaves are compact. \qed
\end{rmk}

\begin{rmk}
By \eqref{e:DeltaT} the condition that $\es$ is in $\calN_T$ occurs if and only if $\Delta T$ is  in the ideal generated by $T$.  One simple, but rather common, case when 
this happens is when $S$ is a solution of the classical master equation $\{S,S\}=0$, which implies 
$T=-\ii\hbar\Delta S$ and hence $\Delta T=0$. This may give the impression that we have an amenable way of treating anomalous theories, 
i.e., theories in which the action $S$ is a solution to the classical master equation that cannot be deformed to a solution of the quantum one. 
The problem, apart from having to consider an algebra of preobservables different from $\calA$, is that it might be difficult to find a natural 
gauge fixing Lagrangian $\calL$ in $\calC_T$. 
\end{rmk}

In the rest of the paper we will specialize to the case of an AKSZ theory where we deform the de~Rham differential in the source manifold to the equivariant differential w.r.t.\ the infinitesimal action of some Lie algebra. In this case, several pleasant facts occur.
First, $\Delta T=0$. Second, there are natural choices of $\calL$ in $\calC_T$. Finally,
we will see that
$\calA_T$ contains an interesting subalgebra, related to the Cartan model, in which $T$ generates again a Lie differential ideal.

\section{Equivariant AKSZ}\label{s:AKSZ}

We now discuss the equivariant extension of the AKSZ construction.
   We follow the standard notations and ideas, e.g. see \cite{Cattaneo:2001ys, roytenberg:courant2} for the review of some basic concepts and notations.
 Let $\Sigma$ be a $d$-dimensional manifold with a Lie algebra $\g$ acting on it via the vector fields 
$v_X$ for any $X\in\g$. Let $\M$ be a graded manifold with a symplectic form of degree $d-1$ and a homological Hamiltonian 
$\Theta\in C^d(\M)$; we denote with $D_\Theta$ its 
Hamiltonian vector field. Let $\F_\Sigma={\Map(T[1]\Sigma,\M)}$ be the AKSZ space of fields. The BV vector field is given by
$$
Q_{BV} = \hat{d_\Sigma} +\hat{D}_\Theta = \{S_{BV},-\},
$$
where $S_{BV}=S_0+S_{\Theta}$ and $S_0$ and $S_\Theta$ are the Hamiltonians of $\hat{d}_\Sigma$  (which is associated with de Rham differential $d_\Sigma$ on $\Sigma$.) and $\hat{D}_\Theta$, respectively.  
Here we denote with $\hat{v}$ the vector field of $\F_\Sigma$ obtained from a vector field $v$ either of the source $T[1]\Sigma$ or of the target $\M$ by composing it with maps.
Since $D^2_\Theta=0$, $Q^2_{BV}=0$ and $S_{BV}$ solves the classical master equation $\{S_{BV},S_{BV}\}=0$.

The space of functionals $\A=C(\F_\Sigma)$ is a $\g$-dg algebra with differential $Q_{BV}=\hat{d}_\Sigma+\hat{Q}$, contraction 
$\hat{\iota}_{v_X}$ and Lie derivative 
$\hat{L}_{v_X}$ for any $X\in\g$. They are all Hamiltonian vector fields with Hamiltonians $S_{BV}, S_{\hat{\iota}_{v_X}}$ and $S_{\hat{L}_{v_X}}$ respectively (see Appendix \ref{appendix_equivariant_cohomology} for notations). 
We recall that $\A[u]=C(\F_\Sigma)\otimes S\g^*$. We denote with $\langle e_a\rangle$ a basis of $\g$.

Let us define the equivariant extension of the BV action in the Cartan model as
\begin{equation}
 \label{BV_action_cartan}
 S_{BV}^{c} = S_{BV} - u^aS_{\hat{\iota}_{v_a}}\;,
\end{equation}
so that 
\begin{equation}\label{BV_cartan_model}
Q_{BV}^{c} = \{S_{BV}^c,-\}= \hat{d}_\Sigma + \hat{D}_\Theta - u^a \hat{\iota}_{v_a}\;
\end{equation}
is the differential of the Cartan model of equivariant cohomology. If for $X\in\g$ we denote $L_X = -X^a f_{ab}^c u^b\frac{\partial\ }{\partial u^c}$ and 
$\L_X=L_X+\hat{L}_{v_X}$,  then we have that
$$
\L_X S_{BV}^c=0\;,
$$
{\it i.e.} $S_{BV}^c\in \A[u]^\g$; moreover $S^c_{BV}$ satisfies the {\it modified Classical Master Equation}
\begin{equation}\label{modCME}
\frac{1}{2}\{S_{BV}^c, S_{BV}^c\} + u^a S_{\hat{L}_{v_a}} =0\;.
\end{equation}


As in (\ref{qme_anomaly}), we define 

\begin{equation}\label{equivariant_CME_cartan}
T := \frac{1}{2}\{S^c_{BV},S^c_{BV}\} -i\hbar\Delta S^c_{BV}= 
- u^a S_{\hat{L}_{v_a}} -i\hbar\Delta S^c_{BV} \;,
\end{equation}
so that
\begin{equation}
 T = - u^a (S_{\hat{L}_{v_a}}+i\hbar \Delta S_{\hat{\iota}_{v_a}})-i\hbar \Delta S_{BV}\;.
\end{equation}

Since $S_0$ and $S_{\hat\iota_{v_a}}$ are quadratic in the fields, then $\Delta$ applied to them will produce constant functionals so that
\begin{equation}\label{Delta_regularization}
\{\Delta S_0,-\}=\{\Delta S_{\hat{\iota}_{v_X}},-\}=0\,,\;\;\;X\in\g\;.
\end{equation}
These functionals should be thought of as regularized traces of the corresponding operators $\hat{d}_\Sigma$ and $\hat{\iota}_{v_X}$; since these operators
are odd a reasonable definition of the trace should be $0$, but it is enough to assume from now on that our regularization of $\Delta$ 
satisfies (\ref{Delta_regularization}) (see the footnote in section \ref{s:QME}).  

Equations (\ref{Delta_regularization}) have the following interesting consequences. The first one is that, consistent with the rules of the BV algebra, 
$\Delta S_{\hat{L}_{v_X}}=0$ for all $X\in\g$; in fact
$$
\Delta S_{\hat{L}_{v_X}}=\Delta\{S_0,S_{\hat{\iota}_{v_X}}\} =\{\Delta S_0,S_{\hat{\iota}_{v_X}}\} \pm \{S_0,\Delta S_{\hat{\iota}_{v_X}}\} = 0\;.
$$
This in particular implies that $\Delta T=0$,
\begin{equation}\label{Delta_L}[\Delta, \hat{L}_{v_X}]=0\;
\end{equation}
and by equation (\ref{e:DeltaT}) also that
\begin{equation}\label{T_equivariantly_closed} 
Q^c_{BV}T=0,
\end{equation}
so that $S^c_{BV}\in {\mathcal N}_T$. We are then in the situation discussed at the end of Remark \ref{rmk_coiso_red}.  
Applying $\Delta$ to $\{S_{\hat{L}_{v_a}},S_{\hat\iota_{v_b}}\}$ we get the relations
\begin{equation}\label{delta_contraction}
f_{ab}^c \Delta S_{\hat\iota_{v_c}} =  0\;.
\end{equation}
The last consequence of (\ref{Delta_regularization}) is that
$$\{T,\calO \}=\{T', \calO \}~,$$
where
$$ T' = - u^a S_{\hat{L}_{v_a}} -i\hbar \Delta S_\Theta\;. $$
Following the general discussion of the previous section, we can now write
$$
\calN_T = \{ \calO \in\calA[u],\ \{T', \calO \}\in\calI_T\}
$$
where $\calI_T$ is the ideal generated by $T$ in $\A[u]$.


We can now define an interesting subalgebra of $\calN_T$. A stronger condition than $\{T', \calO \}\in \calI_T$ is given by the conditions 
\begin{equation}
 \L_a \calO  =0 =\{\Delta S_\Theta,  \calO \}\;\;\;\; \forall a\;.
\end{equation}
In fact the conditions $\L_a \calO  =0$ for all $a$ imply $u^a \{S_{\hat{L}_{v_a}}, \calO \}=u^a \hat{L}_{v_a} \calO=0$ (note that $u^a L_a =0)$.
We then define 
\begin{equation}
\calN_T'=\{\calO \in\A[u],\ \L_X \calO =0 =\{\Delta S_\Theta, \calO \}\; \forall X\in\g\}\subset \calN_T\;.
\end{equation}

Recall that $\DeltaST$ is the twisted BV laplacian defined in (\ref{twisted_BV}).

\begin{Prop}\label{NTprimePoisson}
Under the hypothesis (\ref{Delta_regularization}), $\calN_T'$ is a Poisson subalgebra that is invariant under both $Q^c_{BV}$ and $\DeltaST$. Moreover, $T\in\calN_T'$.
\end{Prop}

{\it Proof}. A direct computation shows that $[\L_X,Q^c_{BV}]=0$, for all $X\in\g$. Moreover, we have that
\begin{eqnarray*}
\{\Delta S_\Theta, S^c_{BV}\} &=& \{\Delta S^c_{BV},S^c_{BV}\} = \{\frac{i}{\hbar} (T + u^a S_{\hat{L}_{v_a}}), S^c_{BV}\} \cr
&=& \frac{i}{\hbar} (Q^c_{BV}(T)+u^a\L_a S^c_{BV})=\frac{i}{\hbar} Q^c_{BV}(T)=0\;, 
\end{eqnarray*}
where we used (\ref{Delta_regularization}) in the first equality, (\ref{equivariant_CME_cartan}) in the second one and (\ref{T_equivariantly_closed}) in the last one. 
We then proved invariance under $Q^c_{BV}$.

Invariance under $\DeltaST$ follows from $iv$) of the two following Lemmas; the last statement follows from $iii$) of those Lemmas. \qed

\begin{Lem}\label{first_two_lemmata} The following relations are valid for all $X\in\g$:
\begin{enumerate}
              \item[$i$)] $[\L_X,\Delta]=0$;
              \item[$ii$)] $[\L_X,Q_{BV}^c]=0$;
              \item[$iii$)] $\L_X(T)=0$;
              \item[$iv$)] $[\L_X,\DeltaST]=0$.
              \end{enumerate}
\end{Lem}
{\it Proof.} Property $i$) follows from (\ref{Delta_L}) and the fact that $\Delta$ clearly commutes with $L_a$. To prove property $ii$), we first observe that $\L_a$ clearly 
commutes with $\hat{Q}$ and $\hat{d}_\Sigma$; moreover
$$
[\L_a,u^b\hat{\iota}_{v_b}] = [\hat{L}_{v_a},u^b\hat{\iota}_{v_b}] + [L_a,u^b\hat{\iota}_{v_b}] = u^b [\hat{L}_{v_a},\hat{\iota}_{v_b}] + [L_a,u^b]\hat{\iota}_{v_b}=0\;.  
$$
To prove property $iii$), we write 
$T = T_0+ C -i\hbar \Delta S_{BV} $, where 
$$T_0=- u^a S_{\hat{L}_{v_a}}$$
and $C$ is the constant functional ${\color{red} +}i\hbar u^a\Delta S_{\hat{\iota}_{v_a}} $. We prove first that $\L_a(T)=0$. Indeed,
$$
\hat{L}_{v_a}(T_0) = - u^b \hat{L}_{v_a}(S_{\hat{L}_{v_b}}) = - u^b f_{ab}^c S_{\hat{L}_{v_c}} = - L_a (T_0)
$$
so that $\L_a(T_0)=0$. Moreover,
$
\hat{L}_{v_a}(C) = 0$ since $C$ is constant and equation (\ref{delta_contraction}) implies $L_a C=0$. Finally
$\hat{L}_{v_a}\Delta S_{BV} = \Delta \hat{L}_{v_a} S_{BV}=0 $ from (\ref{Delta_L}) and obviously $L_a \Delta S_{BV}=0$.

Property $iv$) is an immediate consequence of the previous ones. \qed

\begin{Lem}\label{second_two_lemmata}
Let $V_{\Delta S_\Theta}$ be the Hamiltonian vector field of $\Delta S_\Theta$. We have that
\begin{enumerate}
              \item[$i$)] $[V_{\Delta S_\Theta},\Delta]=0$;
              \item[$ii$)] $[V_{\Delta S_\Theta},Q_{BV}^c]=0$;
              \item[$iii$)] $V_{\Delta S_\Theta}(T)=0$;
              \item[$iv$)] $[V_{\Delta S_\Theta},\DeltaST]=0$.
              \end{enumerate}
\end{Lem}
{\it Proof.} Property $i$) follows since, being $\Delta$ a derivation of the odd bracket, $[\Delta, V_{\Delta S_\Theta}]= V_{\Delta^2 S_\Theta}=0$.  
In order to prove $ii$), let us write
$$
\{\Delta S_\Theta, S^c_{BV}\}=\{\Delta S^c_{BV}, S^c_{BV}\} = \frac{1}{2} \Delta\{S^c_{BV}, S^c_{BV}\} = -u^a\Delta S_{\hat{L}_{v_a}} = 0\;,
$$
where the second equality holds because $\Delta$ is a derivation of the odd bracket, the third follows from the modified classical master equation (\ref{modCME}) and 
the fourth one from (\ref{Delta_L}). 

Let us prove $iii$). From the obvious equation $\{T,T\}=0$ we finally get
$$
0= \{u^a S_{\hat{L}_{v_a}} +i\hbar \Delta S_\Theta, T\} = u^a \L_a(T) + i\hbar\{\Delta S_\Theta, T\} = i\hbar\{\Delta S_\Theta, T\}\;,
$$ 
where we used $iii$) of Lemma \ref{first_two_lemmata}. Property $iv$) is a consequence of $i-iii$). \qed

\medskip

Remark that $Q^c_{BV}$ squares to zero when restricted to $\calN_T'$; we call $(\calN_T',Q^c_{BV})$ the algebra of {\it classical equivariant BV preobservables}. 
A {\it classical equivariant BV observable} is a classical equivariant BV observable that is closed under 
$Q^c_{BV}$.
\smallskip

\begin{Lem}
The ideal $\I_T'$ in $\calN_T'$ generated by $T$ is a $\DeltaST$-invariant Poisson ideal.
\end{Lem}
{\it Proof.}
Let $\calO T\in \I_T'$ and $U\in\calN_T'$. We then compute
$$
\{U,\calO T\}= \{U,\calO \}T \pm \calO \{U,T\}= \{U,\calO \}T\,,
$$
where $\{U,T\}=\{U,T'\}=0$ since $U\in\calN_T'$. Moreover, $\{U,\calO \}\in\calN_T'$ since $U,  \calO \in\calN_T'$ and Proposition \ref{NTprimePoisson}, so that $\{U,\calO T\}\in\I_T'$. We then see that
$$
\DeltaST(\calO T)= (\DeltaST \calO ) T \pm \{\calO,T\}= (\DeltaST \calO )T \,, 
$$
as a consequence of (\ref{DeltaSTderivation}). Finally, as a consequence of points $iv$) of Lemmata \ref{first_two_lemmata} and \ref{second_two_lemmata} we
check that $\DeltaST \calO \in \calN_T'$ so that $\calI_T'$ is $\DeltaST$-invariant. \qed
\smallskip

We define the algebra of {\it quantum equivariant preobservables} as $\A_T'=\calN_T'/I_T'$ with its induced differential:

\begin{equation}
\label{induced_differential}
\DeltaST[\calO ] := [\DeltaST \calO]=\left[\left(\Delta + \frac{i}{\hbar}Q^c_{BV}\right)\calO \right] \;.
\end{equation}
A {\it quantum equivariant observable} is an equivariant preobservable which is $\DeltaST$ closed, for instance the equivalence class of the constant functional.

It is customary to regularize $\Delta S_{\hat\iota_{v_a}}$ and $\Delta S_0$ as zero, see the comment after (\ref{Delta_regularization}). Moreover,
one may also often assume $\Delta S_\Theta =0$ (for instance this is the case for the Poisson Sigma Model with unimodular Poisson structure, see \cite{BZ}). In this case, we have
$$
T = T' = T_0 = -u^a S_{\hat{L}_{v_a}}
$$
and 
$$\calN_T'=\{\calO \in\A[u], \L_X \calO  =0 \; \forall X\in\g\} = \A[u]^\g \;.$$
Remark that the complex $(\A[u]^\g, Q^c_{BV})$ is the Cartan model for the equivariant cohomology of the $\g$-differential algebra $\A=C(\F_\Sigma)$. 
%

Finally, let us discuss gauge fixing when the target manifold $\M$ is a graded vector space $V$, so that the space of BV fields is
$\F_\Sigma=\Omega\Sigma\otimes V$. Let us introduce an invariant metric on $\Sigma$ and let us define $\L=\Omega^{co}(\Sigma)\otimes V$, where 
$\Omega^{co}(\Sigma)$ stands for coexact forms.
In general, due to harmonic forms of $\Sigma$, $\L$ is only isotropic, but let us ignore this issue at the present level of discussion. Since the 
invariance of the metric implies that $[L_{v_X},d^\dagger]=0$, we have that $S_{\hat{L}_{v_X}}|_{\L}=0$. 
The characteristic foliation defined by $\{T,-\}$ coincides with the infinitesimal 
$\g$-action so that we have to require that $G$ is compact (see the discussion in Remark \ref{rmk_coiso_red}). 

 \begin{rmk}
 Some instances of the construction of this paper for an equivariant extension of a BV action with a term that breaks the master equation have appeared before. For example, in \cite{CFequiv} an $S^1$-equivariant version of the Poisson sigma model on a disk is studied; the equivariant extension of the BV action is hinted at in Example 2 and an invariant gauge fixing is behind the choice of propagator of Section 5.3; the whole Feymnan diagram expansion, which is at the core of that paper, is the one corresponding to the equivariant BV theory.  Another example is Geztler's paper \cite{G-covar} where the special case of classical 
 BV-equivariance under source diffeomoprhisms for one-dimensional systems is considered (as this paper only focuses on classical aspects, no discussion of allowed gauge fixings appears there). The present paper includes these two examples, and introduces more (see Sections 4 and 5), in a general conceptual framework.
To the best of our knowledge the first discussion about the relation between BV formalism and equivariant localization can be found in \cite{nersessian}.
  For a recent general overview of the dependency of a BV theory on the gauge-fixing Lagrangian see \cite{Mikhailov:2016rkp}.
 \end{rmk}

\section{Equivariant two dimensional SYM}\label{s:2D}

We discuss here the equivariant extension of two dimensional supersymmetric Yang-Mills theory; we use version of the AKSZ approach developed in \cite{bciz:EquivA-mod}.

Let $\Sigma_2$ be a two dimensional closed manifold\footnote{We may relax this condition if we can guarantee that Stokes theorem works
 by imposing the appropriate boundary conditions or appropriate decay at infinity. This comment is applicable to AKSZ construction in general.} 
 and $\g$ a Lie algebra acting on it. Let us consider the AKSZ theory with target 
$T^*[1](\k[1]\times \k[2])$, where $\k$ is a Lie algebra (not to be confused with $\g$). The index $\alpha$
appearing in the following formulas runs over a basis of $\k$ and $a$ over a basis of $\g$.
If $c,\phi$ are the Lie algebra coordinates of $\k$ of degree $1,2$ respectively and $\xi,\tilde\xi$ 
the momenta of degree $0,-1$ respectively, then the homological Hamiltonian on  $T^*[1](\k[1]\times \k[2])$ reads
\beq\label{eq.ham.}
 \Theta = 
\frac{1}{2}\xi_\alpha [c,c]^\alpha 
  {+} \widetilde{\xi}_\alpha [c,\phi]^\alpha + \xi_\alpha\phi^\alpha	~,
\eeq
so that $D(\cdot)=\{\Theta,\cdot\}$ reads:
\beq\label{target_differential}
 &D c= \phi+\frac{1}{2} [c, c]  ~,\\
 &D \phi = [c, \phi] ~,\\
 &D \xi = [c,\xi] - [\phi,\tilde\xi]  ~,\\
 &D \widetilde{\xi} = \xi +[c,\tilde\xi]   ~.
\eeq

The superfields   are defined as different degree components of a map $T[1] \Sigma_2 \rightarrow T^*[1](\k[1]\times \k[2])$
\beq
 &\c = c + A + \xi^\vee ~, \qquad 				&&\superXi = \xi + A^\vee + c^\vee ~,\\
 &\superPhi = \phi + \psi + \widetilde{\xi}^\vee ~, \qquad 	&& \widetilde{\mathbf{\Xi}} = \widetilde{\xi} + \psi^\vee + \phi^\vee ~,
\eeq
   where we use the same letters for the lowest component of superfields as the coordinates on $T^*[1](\k[1]\times \k[2])$.
The equivariant AKSZ action in the Cartan model is
\beq\label{eq.act}
 S_{BV}^{c} = \int_{T[1]\Sigma_2} 
   \mathbf{\Xi}_\alpha\mathbf{\Phi}^\alpha +  \frac{1}{2}\mathbf{\Xi}_\alpha[\mathbf{c},\mathbf{c}]^\alpha
 + \frac{1}{2}\widetilde{\mathbf{\Xi}}_\alpha[\mathbf{\Phi},\mathbf{c}]^\alpha + \mathbf{\Xi}_\alpha\mathrm{d}_G \mathbf{c}^\alpha + 
 \widetilde{\mathbf{\Xi}}_\alpha\mathrm{d}_G \mathbf{\Phi}^\alpha 
\eeq
where $d_G= d_\Sigma- u^a\iota_{v_a}$ is the equivariant differential. We compute the equivariant extension of the BV differential in the Cartan model as
\beq
 \label{BV_cartan}
& Q_{BV}^{c} (A) = \psi' +d_Ac \cr
 &Q_{BV}^{c} (\psi') = +d_A\phi+[c,\psi'] + u^a\iota_{v_a}F(A)\cr
&Q_{BV}^{c}(\phi) = [c,\phi] -u^a\iota_{v_a}\psi'\cr
&Q_{BV}^{c}(c) = \phi +\frac{1}{2} [c,c] - u^a\iota_{v_a}A\,,\cr
&Q_{BV}^c(H) = [c,H] - [\phi,\tilde{\xi}] + u^a\iota_{v_a} d_A\tilde\xi\,,\cr
&Q_{BV}^c(\tilde\xi) = H+[c,\tilde\xi]\,,
\eeq
where $\psi'= \psi -u^a\iota_{v_a}\xi^\vee$, $H= \xi-u^a\iota_{v_a}\psi^\vee$ and $d_A = d_\Sigma + [A,-]$. 
As explained in Section 5.2 of \cite{bciz:EquivA-mod} for the non equivariant case, after the gauge fixing of the AKSZ model all fields can be  identified with the components of the full $N=2$ vector 
supersymmetric multiplet\footnote{For mathematically friendly review of supersymmetric one can consult \cite{Deligne:1999qp}.}.  
 Here after  the gauge fixing procedure we identify  all fields and  the supercharge of the topologically twisted $N = 2$ supersymmetric gauge theory in the so called $\Omega$-background.  

Let us fix an arbitrary invariant metric on $\Sigma_2$ and assume the standard Hodge decomposition of differential forms on $\Sigma_2$ as a sum of harmonic, exact and co-exact parts\footnote{More precisely, a metric and an orientation define the Hodge $\star$ operator, which in turn allows defining the co-differential $d^\dagger:= - \star d \star $ in even dimensions. 
 A form in the image of $d^\dagger$ is called co-exact. Harmonic forms may be identified with forms lying in the intersection of the kernels of $d$ and $d^\dagger$.}
We consider the standard gauge fixing Lagrangian defined by  selecting the subspace of coexact forms (zero modes given by cohomology can be ignored for what concerns the present discussion). The two-form fields are then put to zero, {\it i.e.} $\xi^\vee=\tilde\xi^\vee=c^\vee=\phi^\vee=0$; in order to fix the one-form fields $A$ and $\psi$ we add two sets of {\it equivariant trivial pairs} $\{\bar c, b\}$ and 
$\{\lambda,\rho\}$, respectively.  

Namely the first one is given by $\lambda,\rho\in\Omega^0(\Sigma_2;\k)$ of ghost number $-2$ and $-1$ respectively with momenta $\lambda^\vee,\rho^\vee\in\Omega^2(\Sigma_2;\k^*)$ of ghost degree~$1$ and~$0$\,. The second one is
given by $\bar c, b\in\Omega^0(\Sigma_2,\k)$ of degree $-1,0$ respectively, with momenta $\bar c^\vee,b^\vee\in\Omega^2(\Sigma_2,\k^*)$ of degree $0,-1$.

The gauge fixing fermion is defined as
$$
\Psi = \int_{T[1]\Sigma_2} \lambda\ d_\Sigma \star\psi + \bar c\  d_\Sigma\star A~,
$$
   where $\star$ is Hodge star and  $A$ is  a one form, since we expand around the zero connection. 
 In general  we have to fix a background connection and 
    analyse the expansion around this connection. But in this paper we consider only trivial background.

The Lie algebra $\g$ acts in the direction of the trivial pair with Hamiltonians
$$
S^{tr}_{\hat\iota_{v_a}}= \int_{T[1]\Sigma_2} \rho^\vee_\alpha L_{v_a}\lambda^\alpha+
b^\vee_\alpha L_{v_a}\bar c^\alpha \,,$$
$$S^{tr}_{\hat{L}_{v_a}}=\int_{T[1]\Sigma_2} \lambda^\vee_\alpha L_{v_a}\lambda^\alpha + 
\rho^\vee_\alpha L_{v_a}\rho^\alpha +  \bar c^\vee_\alpha L_{v_a}\bar c^\alpha + 
b^\vee_\alpha L_{v_a}b^\alpha \;.$$ 
The BV action (\ref{eq.act}) will be shifted by a term 
$$ S^{c}_{BV}=\int_{T[1]\Sigma_2} \lambda^\vee_\alpha \rho^\alpha+\bar c^\vee_\alpha b^\alpha - 
u^a\int_{T[1]\Sigma_2} \rho^\vee_\alpha L_{v_a}\lambda^\alpha +b^\vee_\alpha L_{v_a}\bar c^\alpha\;$$
and the full BV action will be 
$$ S^{\rm full}_{BV} = S_{BV} +  S_{BV}^{c}$$
Next we have to guarantee that our gauge fixing corresponds to $T$-Lagrangian. For this we need to check that 
$$\{ S_{BV}, S_{BV} \} =0$$
on our Lagrangian. Some terms vanish identically and for some terms we need to use integration by parts with $L_{v_a}$
 (this is true only when we use invariant metric). 

The BV transformations of the trivial pair $(\lambda,\rho)$ then read 
\begin{eqnarray}
 \label{BV_trivial_pair}
 Q_{BV}^c (\lambda) &=& \zeta + [c,\lambda] ~,\\
 Q_{BV}^c (\zeta) &=& -[\phi,\lambda] + [c,\zeta] + u^a\iota_{v_a} d_A\lambda \;,\nonumber
\end{eqnarray}
where $\zeta= \rho - [c,\lambda]$. By a direct comparison, one can check that $Q^c_{BV}$ restricted to the multiplet $\{A,\phi,\psi',H,\tilde\xi,\lambda,\zeta\}$ acts as 
\[
 Q_{\mathrm{BV}}^c = \delta_{\mathrm{BRST}} + \delta_{\mathrm{susy}} ~,
\]
where $\delta_{\mathrm{susy}}$ is the supercharge in \cite{nekrasov09} and $\delta_{\mathrm{BRST}}$ the usual BRST operator.

\begin{rmk}
In \cite{bciz:EquivA-mod} the susy multiplet was recovered with a slightly different procedure. Indeed, the trivial pair $(\lambda,\rho)$ appeared with an ad hoc procedure, without performing the actual gauge fixing. In this way we missed the fact the they appear in the standard gauge fixing procedure
as the antighost and Lagrange multiplier needed for imposing the gauge fixing condition $d\star\psi=0$.
\end{rmk}
Let us now discuss the classical equivariant BV observables. Following \cite{bciz:EquivA-mod} we look for a map $\ev:\F_{\Sigma_2}\otimes T[1]\Sigma_2\rightarrow T[1]\k[1]$
such that for each $\omega\in C(T[1]\k[1])$
\begin{equation}\label{eq_chain_map}
(Q_{BV}^{c}-d_\Sigma + u^a\iota_{v_a}) \ev^* \omega = \ev^* D\omega.
\end{equation}
A straightforwad computation shows that 
\begin{equation}
\label{Maxim_observables}
\ev^*(c) = c+A\;, \ev^*(\phi) = \phi+\psi' -F(A)
\end{equation}
satisfies (\ref{eq_chain_map}). From (\ref{Maxim_observables}) we see that 
$$
(\hat{L}_{v_a}+L_a - L_{v_a})\ev^*\omega =0\;.
$$

Let now $D\omega=0$ and let $\gamma[u]\in(\Cu\otimes S\g^*)^\g$ be an equivariant cycle as discussed at the end of Appendix \ref{equivariant_cohomology}; we define $\calO^\gamma_\omega \equiv \int_{\gamma[u]}\ev^*\omega$. 
From (\ref{eq_chain_map}) we see that
$$
Q_{BV}^{c}\calO_\omega^\gamma =  \int_{\gamma[u]} (d_\Sigma - u^a\iota_{v_a})\ev^*\omega = \int_{\partial_G\gamma[u]}\ev^*\omega =0\;.
$$
Moreover, from (\ref{Maxim_observables}) it follows that
$$
\L_a \calO_\omega^\gamma= (\hat{L}_{v_a}+L_a)\calO^\gamma_\omega = -\int_{\gamma[u]}L_{v_a}\ev^*\omega =\int_{L_{v_a}\gamma[u]}\ev^*\omega = 0\\,
$$
so that $O^\gamma_\omega\in\A[u]^\g$ is an equivariant classical BV observable.  


\section{Equivariant Donaldson-Witten theory}\label{s:4D}

We analyze here the AKSZ approach to Donaldson-Witten theory \cite{w88}. 
We start with a discussion of the non equivariant case. Our derivation will differ from \cite{Ik} in the gauge fixing procedure.

\subsection{DW from AKSZ}
Let $\k$ be a Lie algebra. In the previous section we could use the Weil model $W(\k)$ as the target of a 2d AKSZ model only after embedding it in the bigger dGA (\ref{target_differential}) that has a natural symplectic form of degree $1$. 
If $\k$ admits an invariant non degenerate symmetric pairing  $\langle~,~\rangle$, then $W(\k)$ admits a natural symplectic form of degree $3$ and can be used as a target of a 4d AKSZ theory.
 
Indeed the graded vector space 
 \bea
  \k[1] \oplus \k[2]
 \eea
 is  equipped with the symplectic structure of degree $3$ 
  \bea
   \omega = \langle \delta c, \delta \phi \rangle~,
  \eea
  where $c$ is the coordinate of degree $1$ and $\phi$ is the coordinate of degree 2. The Hamiltonian function of 
   degree $4$
   \bea
    \Theta = \frac{1}{2} \langle \phi, \phi \rangle + \frac{1}{2} \langle \phi, [c,c]\rangle~,
   \eea
has the Weil differential as Hamiltonian vector field
 \beq
& d_W c = \phi + \frac{1}{2} [c, c]~,\\
 &d_W \phi =[c, \phi]~.
 \eeq
Let us consider now the 4D AKSZ model with this target and source the four dimensional manifold $\Sigma_4$. 
The superfields  are defined as map $T[1]\Sigma_4 \rightarrow  \k[1] \oplus \k[2]$
   \beq
  &\mathbf{c} = c + A + \chi +\psi^\vee + \phi^\vee ~,\\
  &\mathbf{\Phi} = \phi + \psi + \chi^\vee  + A^\vee +  c^\vee~,
  \eeq
    where we use the same letter for lowest component as for the coordinates on $\k[1] \oplus \k[2]$.
The BV symplectic form is
  \bea
   \omega_{BV} = \int_{T[1]\Sigma_4} d^4x d^4\theta~ \langle \delta \mathbf{c} , \delta \mathbf{\Phi} \rangle \label{fullBV-sympl}
  \eea
 and the AKSZ action is
  \bea\label{DW_AKSZ}
   S_{BV} = \int_{T[1]\Sigma_4} \left ( \langle\mathbf{\Phi},  d_\Sigma \mathbf{c} \rangle + \frac{1}{2} \langle \mathbf{\Phi}, \mathbf{\Phi} \rangle 
   + \frac{1}{2} \langle \mathbf{\Phi}, [\mathbf{c},\mathbf{c}]\rangle \right ).\label{fullBV-action}
  \eea
  In terms of the components, the BV symplectic structure can be written as follows
  \begin{eqnarray}
    \omega_{BV} &=& \int\limits_{\Sigma_4} \left ( \delta c \wedge \delta c^\vee + \delta A \wedge \delta A^\vee + 
    \delta \chi \wedge \delta \chi^{\vee} 
+  \delta \psi^\vee \wedge \delta \psi+\delta \phi^\vee \wedge \delta \phi \right ).\nn
  \end{eqnarray}
The BV action in components reads
    \bea
 S_{BV}& =& \int_{\Sigma_4} \Big ( \langle \psi , d_A \chi \rangle  + \frac{1}{2}  \langle \phi, [\chi, \chi] \rangle   +  
 \langle \psi^\vee, ( d_A \phi + [c, \psi])  \rangle +\nn \\
&&  \langle \chi^\vee, ( F  +   [c, \chi]  ) \rangle  +
 + \langle A^\vee , (\psi + d_A c ) \rangle + \langle \phi^\vee , [c, \phi] \rangle \nn \\
&&         +     \langle c^\vee, (\phi + \frac{1}{2}[c, c] )\rangle  +
     \frac{1}{2} \langle \chi^\vee, \chi^\vee \rangle  \Big ) \label{BVDW-action1}
    \eea

\begin{rmk}
The linear terms in anti-fields give us familiar BRST-transformations of the fields. The quadratic term in anti-fields is telling us that they close only on-shell. 
It can be fixed by introducing the couple of two forms $(H, H^\vee)$, even and odd correspondently with $\deg H=0$ and $\deg H^\vee =-1$. The last quadratic 
      term in the action can be replaced as follows
      \bea
        \frac{1}{2} \langle \chi^{\vee}, \chi^\vee \rangle ~\rightarrow ~ -\langle \chi^\vee, H \rangle - \frac{1}{2} \langle H, H \rangle~.
      \eea
       In this way we get an action linear in anti-fields and this is just  canonical embedding of DW BRST-transformations into BV. 
\end{rmk}

   The BV transformations on the superfields are 
   \beq
  &  Q_{BV} \mathbf{c} = d_\Sigma\mathbf{c} + \mathbf{\Phi} + \frac{1}{2} [\mathbf{c}, \mathbf{c}]~,\\
   & Q_{BV} \mathbf{\Phi} = d_\Sigma \mathbf{\Phi} + [\mathbf{c}, \mathbf{\Phi}]~,
   \eeq
   and on the components become
   \beq\label{BV_DW}
   & Q_{BV} c = \phi + \frac{1}{2} [c,c]~,\\
   & Q_{BV} A = \psi + d_A c~,\\
   & Q_{BV} \chi = \chi^\vee + F(A) + [c, \chi]~,\\
   & Q_{BV} \psi^\vee  = d_A \chi + A^\vee + [c, \psi^\vee]~,  \\
   & Q_{BV} \phi^\vee = d_A \psi^\vee + c^\vee + [c, \phi^\vee]~,  \\
   & Q_{BV} \phi = [c, \phi]~,\\
   & Q_{BV} \psi = d_A \phi + [c, \psi]~,\\
   & Q_{BV} \chi^\vee = d_A \psi + [c, \chi^\vee] + [\chi, \phi]~,\\
   & Q_{BV} A^\vee = d_A \chi^\vee + [c, A^\vee] + [\psi^\vee, \phi] + [\chi, \psi]~,\\
   & Q_{BV} c^\vee = d_A A^\vee + [c, c^\vee] + [\phi^\vee, \phi] + [\psi^\vee, \psi] + [\chi , \chi^\vee]~,
   \eeq
   where $d_A = d_\Sigma + [A,~]$ and $F(A)= d_\Sigma A + \frac{1}{2} [A, A]$.

  Let us now discuss the gauge fixing.  Let us introduce a metric on $\Sigma_4$; we split  the two-forms into self-dual and anti-self-dual components
   \bea
   && \chi = \chi^+ + \chi^-~,\nn\\
   && \chi^\vee = \chi^{+\vee} + \chi^{-\vee} ~.\label{decomp-fields}
   \eea
As gauge fixing we impose $\chi^-=0$ and $\chi^{\vee +} =0$ and we require that the other forms are co-exact  (as before we use the Hodge decomposition to define our Lagrangian).
Co-exactness in sectors $(c, c^\vee)$ and $(\phi, \phi^\vee)$ implies 
      that $c^\vee=0$ and $\phi^\vee =0$. To impose the co-exactness in sectors $(A, A^\vee)$ and $(\psi, \psi^\vee)$ we use the standard procedure of introducing extra trivial sectors. To impose the co-exactness on $(A, A^\vee)$ we introduce zero forms $(\bar{c}, b)$ with $\deg \bar{c} =-1$, $\deg b =0$ and  their antifields $(\bar{c}^\vee, b^\vee)$ which are top forms with $\deg \bar{c}^\vee =0$ and $\deg b^\vee = -1$. To the BV action (\ref{DW_AKSZ}) we can 
     add the following terms of degree zero
     \bea\label{trivial1}
  S_{tr,1}= \int \Big ( \langle  \bar{c}^\vee,  ( b + [c, \bar{c}] ) \rangle + \langle b^\vee,  ( [c,b] - [\phi, \bar{c}] ) \rangle \Big )~. 
     \eea
With this choice, $b$ and $\bar c$ transform correctly under the gauge transformations.  It is easy to check that the standard BV trival pair is recovered by a simple field redefinition.  Now we have to repeat the same trick for $(\psi, \psi^\vee)$ sector. Let us introduce zero forms $(\varphi, \eta)$ with degrees $\deg \varphi = -2$, $\deg \eta = -1$ 
        and their top forms anti-fields $(\varphi^\vee, \eta^\vee)$
        with $\deg \varphi^\vee = 1$, $\deg \eta^\vee = 0$. To the BV action we add the following term of degree zero
        \bea\label{trivial2}
      S_{tr,2} = \int \Big ( \langle \varphi^\vee, ( \eta + [c, \varphi] )\rangle  +   \langle \eta^\vee , ( [c, \eta] + [\varphi, \phi] ) \rangle \Big )
                \eea
Finally
$$
S_{BV}'=S_{BV}+ S_{tr,1} + S_{tr,2}
$$
satisfies the master equation.  The gauge fixing fermions will be 
     \bea\label{fermionic_gf}      {\bf \Psi} = \int \langle \bar{c}, d \star A \rangle+ \int \langle \varphi, d \star \psi \rangle~,
        \eea
    where $A$ is assumed to be a one-form and we expand around zero connection.

After the gauge fixing we get the following residual gauge transformations
    \bea
   && \delta c = \phi + \frac{1}{2} [c,c]~,\\
   && \delta A = \psi + d_A c~,\nn\\
   && \delta \chi^+ = F^+ + [c, \chi^+]~,\nn\\
   && \delta \phi = [c, \phi]~,\nn\\
   && \delta \psi = d_A \phi + [c, \psi]~,\nn\\
   && \delta \chi^{-\vee} = d_A \psi + [c, \chi^{-\vee}] ~,\nn\\
   && \delta \bar{c} =  b + [c, \bar{c}] ~,\nn\\
   && \delta b =  [c,b] - [\phi, \bar{c}] ~,\nn\\
   && \delta \varphi = \eta + [c, \varphi]~,\nn\\
   && \delta \eta= [c, \eta] + [\varphi, \phi]~.\nn 
   \eea
    Here the bosonic field $\chi^{-\vee}$ is auxilary field of degree $0$ and it can be integrated out. The above transformations square to zero except for $\chi^+$
    \bea
     \delta^2 \chi^+ = (d_A \psi)^+ + [\phi, \chi^+]~,
    \eea
     which is equation of motion. Actually this is easily seen from the full BV action 
    (\ref{BVDW-action1}) which has linear and quadratic terms in anti-fields. 
It is clear that the multiplet $(A,\phi,\psi,\chi^+,\varphi,\eta)$ reproduces the vector multiplet
appearing in \cite{w88}

 \subsection{Equivariant DW theory}
 
Let us now consider that the Lie algebra $\g$ acts on $\Sigma_4$ with vector fields $v_X$, $X\in\g$.
The equivariant extension is obtained by replacing $d_\Sigma$ by $d_G = d_\Sigma - u^a\iota_{v_a}$ in the AKSZ action.
  \begin{eqnarray}
   S_{BV}^c &=& \int_{T[1]\Sigma_4}\left ( \langle\mathbf{\Phi},  d_G \mathbf{c} \rangle + \frac{1}{2} \langle \mathbf{\Phi}, \mathbf{\Phi} \rangle 
   + \frac{1}{2} \langle \mathbf{\Phi}, [\mathbf{c},\mathbf{c}]\rangle \right )~.\label{fullBV-action-equiv}\cr
&=&S_{BV} -u^a \int \Big ( \psi \iota_{v_a} \phi^\vee + \chi^\vee \iota_{v_a} \psi^\vee + A^\vee \iota_{v_a} \chi + c^\vee\iota_{v_a}A\Big )\;.
  \end{eqnarray}
   The BV transformations are 
   \bea
  &&  Q_{BV}^c \mathbf{c} = d_G\mathbf{c} + \mathbf{\Phi} + \frac{1}{2} [\mathbf{c}, \mathbf{c}]~,\\
   && Q_{BV}^c \mathbf{\Phi} = d_G \mathbf{\Phi} + [\mathbf{c}, \mathbf{\Phi}]~.
   \eea
 which in components are written as (we list only relevant fields)
    \bea
   && Q_{BV}^c c = \phi  + \frac{1}{2} [c,c]- u^a\iota_{v_a} A~,\\
   && Q_{BV}^c A =  \psi + d_A c - u^a\iota_{v_a} \chi~,\nn\\
   && Q_{BV}^c \chi = \chi^\vee + F + [c, \chi]- u^a\iota_{v_a} \psi^\vee ~,\nn\\
   && Q_{BV}^c \phi = [c, \phi]-u^a\iota_{v_a} \psi ~,\nn\\
   && Q_{BV}^c \psi = d_A \phi + [c, \psi]-u^a\iota_{v_a} \chi^\vee~,\nn
   \eea
We should stress that in writing A we are expanding round the trivial connection. Round an instanton background the formula will be modified.
 
The procedure of gauge fixing can be done in the same way of the non equivariant case, provided we choose an invariant metric on $\Sigma_4$. The solution of the equivariant master equation is now
 \bea
S_{BV}^{c'}=  S_{BV}^c +  \int \Big ( \langle  \bar{c}^\vee,  ( b + [c, \bar{c}] ) \rangle + \langle b^\vee,  (L_v \bar{c} +  [c,b] - [\phi, \bar{c}] ) \rangle \Big ) \nn \\
  +   \int \Big ( \langle \varphi^\vee, ( \eta + [c, \varphi] )\rangle  +   \langle \eta^\vee , (L_v \varphi+  [c, \eta] + [\varphi, \phi] ) \rangle \Big )\;.
  \eea
 As before the we impose $\chi^-=0$ and $\chi^{\vee +} =0$ and for the other fields we choose the fermonic gauge fixing (\ref{fermionic_gf}).
  As before we have to check that we deal with $T$-Lagrangian gauge fixing and this is guaranteed by choosing the invariant metric
   for the gauge fixing. 
  
 To match with the standard gauge theory we need to do some field redefinitions, e.g. $\tilde{\psi} =  \iota_v \chi +   \psi$ etc.
  One can immediately  recognize the transformations for the equivariant Donaldson-Witten theory (also known as topologically 
   twisted  $N=2$ supersymmetric gauge theory in $\Omega$-background)  \cite{nekrasov03, nkerasov-ok}. 
   To state the relation between our formalism and the $\Omega$-background or equivariant Donaldson-Witten theory, one can take $v^a$ to be the two rotations of $\mathbb{C}^2$ and evaluate $u^a$ at  $\epsilon^a$. Assuming that evaluating $u^a$ at $\epsilon^a$ commutes with the path integral (provided $\epsilon^a$ are not in the support of the equivariant cohomology class produced by the path integral), we will recover the $\Omega$-background calculation. Our assumption about the commutativity is only demonstrated for finite dimension compact manifolds, see section 5 of  
  \cite{AB-moment}. It is beyond the scope of this paper to address this issue for infinite dimension. We will also leave for a future paper the question of utilising our equivariant BV framework to treat ghost zero modes and problems the like.

 Let us point out that the present 4D equivarint AKSZ construction can be generalized in different directions. For example we can adopt 
  alternative decomposition (\ref{decomp-fields}) of the two-forms into self-dual and anti-self dual parts following ideas presented in
  \cite{festuccia}. This will lead to alternative gauge fixing and the resulting theory corresponds to cohomological theory which appeared in
  the Pestun's localization calculation \cite{Pestun} (see  \cite{festuccia} for the corresponding cohomological description).

\appendix

\section{Equivariant cohomology and homology}\label{equivariant_cohomology}\label{appendix_equivariant_cohomology}
Let $\g$ be a Lie algebra. A $\g$-differential algebra $\A$ is a differential graded algebra $(\A,d)$ with $L_X\in\Der^0\A$ and $\iota_X\in\Der^{-1}\A$,
depending linearly on $X\in\g$ and satisfying the rules of Cartan's calculus:
$$
[L_X,d]=0\,,\;\; [L_X,L_Y]=L_{[X,Y]}\,,\; [\iota_X,d]=L_X\,,\;[\iota_X,\iota_Y]=0\,,\; [\iota_X,L_Y] =\iota_{[X,Y]}
$$
The basic subalgebra is defined as $\A_{bas}=\{a\in \A,\ L_Xa=\iota_X a=0,\ \forall\ X\in\g\}$.

Let $\{t_a\}$ be a basis of $\g$. We denote with $W(\g)=(\Lambda\g^*\otimes S\g^*,d_W)$ the Weil $\g$-differential algebra defined as 
\begin{eqnarray}
\dW \theta^a &=& u^a  + \frac{1}{2} [\theta,\theta]^a \\
\dW u^a &=& [\theta,u]^a
\end{eqnarray}
where $\deg \theta=1$ and $\deg u=2$ and $\iota_a=\frac{\partial\ }{\partial \theta^a}$ and $L_a = \{\iota_a,d_W\}$.
The complex $(W(\g),\dW)$ is acyclic; the basic subcomplex is given by $\C[u]^\g=S(\g^*)^{\g}$  the invariant polynomials in $u$ with
  the restriction of $\dW$ to the basic subcomplex.

We can define on $\A\otimes W(\g)$ the obvious tensor product structure of $\g$-differential algebra. We denote with $H_G(\A)$ the cohomology of the basic subcomplex $\A_G=(\A\otimes W(\g))_{basic}$, that is called {\it the Weil model} for $H_G(\A)$. 

We can also consider the graded algebra $\A[u]=\A\otimes S(\g^*)$ equipped with $d_G=d-u^a\iota_{v_a}$ and the diagonal $\g$-action. Since $d_G^2= u^aL_{v_a}$, then $(\A[u]^{\g},d_{G})$ where 
$\A[u]^\g=\{A\in\A[u]\ |\ L_XA=0,\ \forall X\in\g\}$, is a dg algebra. We call it {\it the Cartan model} for $H_G(\A)$. 
In order to prove that its cohomology is isomorphic to $H_G(\A)$ it is enough to check that
\begin{equation}\label{weil_cartan_isomorphism}
I = \exp{[-(\iota_{v_a}\otimes \theta^a)]}: \A\otimes W(\g)\rightarrow \A\otimes W(\g)
\end{equation}
restricts to an isomorphism of $dg$-algebras
$I:\A_G\rightarrow (\A[u])^{\g}$.

Let $\g$ act on the smooth manifold $\Sigma$; the dg algebra of forms $(\Omega(\Sigma),d_\Sigma)$ is a $\g$-differential algebra
with $L_{v_X},\iota_{v_X}$ being the Lie derivative and contraction by the fundamental vector field $v_X$ of $X\in\g$. We denote with $H_G(\Sigma)$ the $G$-equivariant cohomology. In particular we have that $H_G(*)= S(\g^*)^{\g}$.

Let $(\Cu^\bullet(\Sigma),\partial)$ denote the complex of de Rham currents where $\Cu^k(\Sigma)= (\Omega^{n-k}(\Sigma))^*$ and the differential is defined by duality. 
By duality $(\Cu^\bullet(\Sigma),\partial)$ inherits the structure of $\g$-differential algebra. We can then define the Cartan model
$(\Cu(\Sigma)\otimes S\g^*)^\g$ and the Weil model $(\Cu(\Sigma)\otimes W(\g))_{basic}$; we call their cohomology the {\it equivariant homology} of $\Sigma$.

\end{document}